\documentclass[fleqn,twoside]{article}
\usepackage{espcrc2}
\usepackage{longtable}
\def\arcmin{\hbox{$^\prime$}}
\usepackage{graphicx}
\usepackage{epsfig}

\def\lum{erg~s$^{-1}$}

\title{BeppoSAX WFC monitoring of the Galactic Center region}

\author{Jean~in~'t~Zand\address[SRON]{SRON National Institute for
        Space Research, Sorbonnelaan 2, 3584 CA Utrecht, the Netherlands}
        \address[SIU]{Astronomical Institute, Utrecht University,
        P.O. Box 80000, 3508 TA Utrecht, the Netherlands},
        Frank Verbunt\addressmark[SIU],
        John Heise\addressmark[SRON]\addressmark[SIU],
        Angela Bazzano\address[IAS]{CNR Istituto di Astrofisica Spaziale e Fisica Cosmica, Via de Fosso del Cavaliere, 00133 Roma, Italy},
        Massimo Cocchi\addressmark[IAS],
        Remon Cornelisse\address[SOUTH]{Dept. of Physics and Astronomy, University of Southampton, Hampshire SO17 1BJ, U.K.},
        Erik Kuulkers\address[ESTEC]{ESA-ESTEC, SCI-SDG, Keplerlaan 1, 2201 AZ Noordwijk, the Netherlands},
        Lorenzo Natalucci\addressmark[IAS],
        Pietro Ubertini\addressmark[IAS]
}
       
\begin{document}

\begin{abstract}
We review the results obtained with the Galactic center campaigns of
the BeppoSAX Wide Field X-ray Cameras (WFCs). This pertains to the
study of luminous low-mass X-ray binaries (LMXBs). When pointed at the
Galactic center, the WFC field of view contains more than half of the
Galactic LMXB population. The results exemplify the excellent WFC
capability to detect brief X-ray transients. Firstly, the WFCs
expanded the known population of Galactic thermonuclear X-ray bursters
by 50\%. At least half of all LMXBs are now established to burst and,
thus, to contain a neutron star as compact accretor rather than a
black hole candidate. We provide a complete list of all 76 currently
known bursters, including the new case 1RXS
J170854.4-321857. Secondly, the WFCs have uncovered a population of
weak transients with peak luminosities up to $\sim10^{37}$~\lum\ and
durations from days to weeks. One is the first accretion-powered
millisecond pulsar SAX~J1808.4-3658. Thirdly, the WFCs contributed
considerably towards establishing that nearly all (12 out of 13)
luminous low-mass X-ray binaries in Galactic globular clusters contain
neutron stars rather than black holes. Thus, the neutron star to black
hole ratio in clusters differs from that in the Galactic disk at a
marginal confidence level of 97\%.
\end{abstract}

\maketitle
\renewcommand{\topfraction}{1.}
\renewcommand{\bottomfraction}{1.}

\section{INTRODUCTION}

Bleeker \cite{bl03} reviewed in general terms the prospects and
results of the BeppoSAX Wide Field Camera instrument package
(``WFCs''). The unique capability of matching a wide field of view of
$40^{\rm o}\times40^{\rm o}$ per each of two identical cameras with a
good angular resolution of 5\arcmin\ \cite{jag97} not only led to a
revolution of gamma-ray burst research (e.g., \cite{pir03,hei03}) but
also to a serious advance of our knowledge on X-ray bursts and other
transient emission processes in low-mass X-ray binaries (LMXBs). Much
of the relevant data was acquired during the only dedicated BeppoSAX
observation program involving the WFCs as prime instrument. This
encompasses observations pointed at the Galactic center during
semi-yearly visibility windows that lasted from mid February to mid
April and from mid August to mid October. Table~\ref{tab1} summarizes
the twelve campaigns during the six-year BeppoSAX lifetime. The
combined exposure time represents 8\% of the total BeppoSAX exposure
budget. The success of this program may be anticipated by the mere
fact that the field of view of one WFC emcompasses more than half the
Galactic LMXB population according to pre-WFC catalogs
\cite{par95}. Another important ingredient for the success is that
nearly all observations were analyzed in a near to real-time
fashion. This was possible thanks to the 24 hr per day, 7 days a week,
manning of the BeppoSAX Science Operations center (which was also
crucial to the success of the GRB program of BeppoSAX \cite{pir03})
and the dedicated support by the 'duty scientists'.

Tied to this program a dedicated target-of-opportunity program was in
place for the BeppoSAX Narrow Field Instruments (NFIs) to follow up
new transients or bursters. This program was triggered 14 times, with
exposure times between 20 and 40 ksec. Whenever new transients or
bursters were discovered, these were announced in IAU circulars. This
happened on 20 occasions and triggered independent TOO programs on
BeppoSAX, the Rossi X-ray Timing Explorer (RXTE), XMM-Newton, and
ground-based radio and optical telescopes, illustrating a community
service of the WFC program.

\begin{table}[!t]
\caption{WFC observation campaigns on the Galactic center. The
effective exposure times are for the position of the Galactic
center. Adapted from \cite{ha03,we03}.\label{tab1}}
\begin{tabular}{ccrr}
\hline 
\multicolumn{2}{c}{Campaign} & 
\multicolumn{1}{r}{\# obs.} & 
\multicolumn{1}{r}{$t_{\mathrm{exp}}\,(\rm{ks})$} \\
\hline
1996 & Aug.15--Oct.29 & 67 & 1017   \\
1997 & Mar.02--Apr.26 & 21 &  654   \\
1997 & Sep.06--Oct.12 & 13 &  302   \\
1998 & Feb.11--Apr.11 & 17 &  551   \\
1998 & Aug.22--Oct.23 & 10 &  410   \\
1999 & Feb.14--Apr.11 & 14 &  470   \\
1999 & Aug.24--Oct.17 & 24 &  801   \\
2000 & Feb.18--Apr.07 & 21 &  633   \\
2000 & Aug.22--Oct.16 & 29 &  767   \\
2001 & Feb.14--Apr.23 &  5 &  215   \\
2001 & Sep.04--Sep.30 &  7 &  284   \\
2002 & Mar.05--Apr.15 &  5 &   91   \\
\hline
\end{tabular}
\end{table}

In the present paper, we provide a general overview of the
accomplishments obtained through the WFC Galactic center program. We
categorize the achievements in three areas: thermonuclear X-ray
bursts, transients, and luminous globular cluster sources.

\section{THERMONUCLEAR \\ X-RAY BURSTS}

\begin{figure}[t]
\includegraphics[width=\columnwidth]{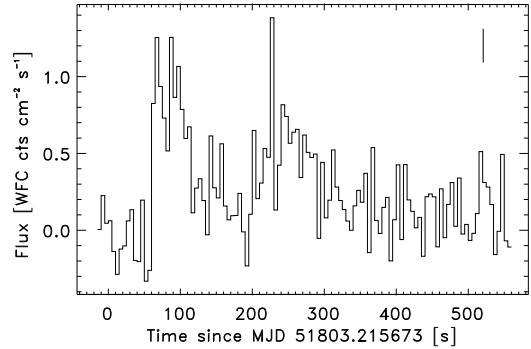}
\caption[]{WFC 2-28 keV light curve of burst from 1RXS
J170854.4-321857.\label{figlc1708}}
\end{figure}

Arguably, the most interesting results obtained are those on type-I
X-ray bursts.  These are events that last between a few seconds and a
few minutes, exhibit a much faster rise than (usually exponential)
decay, and radiate X-ray spectra having simple black body shapes with
temperatures up to roughly 3 keV that cool during decay. The origin of
these bursts lies in thermonuclear flashes in the top layers of a
neutron star that consist of freshly accreted material from a usually
non-degenerate companion star in a close orbit \cite{woo76,mar77}. The
duration of an X-ray burst is mainly determined by the hydrogen
abundance in the layer that is unstably burned during the flash;
longer X-ray bursts indicate larger hydrogen abundances. For reviews
we refer to \cite{le93,st03,cum03}.

The WFC observations resulted in roughly 2200 type-I X-ray burst
detections (1800 from the Galactic center field; see also \cite{ku03}
and \cite{cor03b})\footnote{Since the search for bursts is not
complete yet, these totals could change by 5\%} from 54 sources,
and have increased the X-ray burster population by roughly 50\% from
the pre-WFC era \cite{par95}. In Table~\ref{tab2} we provide a
complete list of the 76 currently known Galactic X-ray bursters. This
represents half the total known LMXB population \cite{liu01}. The 23
bursters not seen bursting with the WFCs are either known to be
transients not active during the WFC era or very sporadic bursters
(e.g., Cir X-1 \cite{ten86a,ten86b,saz03}).

\begin{table*}[!t]
\caption[]{List of all 76 Galactic type-I X-ray bursters currently known.
Updated up to November 2003\label{tab2}. References are to WFC publications.}

\begin{tabular}{ll|ll|ll}
\hline
Source           & Ref. & Source & Ref. & Source & Ref.\\
\hline

MX 0513-40 (NGC 1851)      & \cite{kuu02b} & 1E 1724-3045 (Terzan 2)          & \cite{kuu02b,coc99b} & 2S 1803-245$^{\rm w}$            & \cite{mul98}  \\
4U 0614+09                 & 	           & GX 354-0                         & \cite{cor03}         & SAX J1806.5-2215$^{\rm w}$       & \cite{zan98a} \\
EXO 0748-676               & 	           & KS 1731-260$^{\rm s}$            & \cite{cor03,kuu02}         & SAX J1808.4-3658$^{\rm w}$       & \cite{zan01}  \\
GS 0836-429                & 		   & XB 1733-30 (Terzan 1)            &                      & SAX J1810.6-2609$^{\rm w}$       & \cite{nat01}  \\
2S 0918-549                & \cite{cor02a} & Rapid burster (Liller 1)         &                      & XTE J1814-337                    &               \\
4U 1246-588$^{\rm w}$      & \cite{pir97}  & SLX 1735-269$^{\rm w}$           & \cite{baz97}         & GX 13+1                          &               \\  
4U 1254-69$^{\rm s}$       & \cite{zan03a} & 4U 1735-44$^{\rm s}$             & \cite{cor00}         & 4U 1812-12                       & \cite{coc00}  \\
4U 1323-62                 & 	           & SLX 1737-282$^{\rm w}$           & \cite{zan02a}        & GX 17+2                          &               \\
SAX J1324.5-6313$^{\rm w}$ & \cite{cor02a} & GRS 1741.9-2853$^{\rm w}$        & \cite{coc99a}        & SAX J1818.7+1424$^{\rm w,a}$     & \cite{cor02a} \\
Cen X-4$^{\rm x}$          & 	           & KS 1741-293                      & \cite{zan90,zan98c}  & 4U 1820-303$^{\rm s}$ (NGC 6624) & \cite{kuu02b,cor03} \\
Cir X-1                    & 	           & A 1742-294                       & \cite{cor03}         & AX J18245-2451  (M28)            &               \\
4U 1608-522                & \cite{smi99}  & A 1742-289                       &                      & GS 1826-24$^{\rm w}$             & \cite{cor03,ube99,coc99d}  \\  
4U 1636-536$^{\rm s}$      & \cite{cor03}	           & A1744-361                        &                      & SAX J1828.5-1037$^{\rm w}$     & \cite{cor02a} \\
MXB 1658-298               & \cite{zan99a} & SLX 1744-299                     &                      & 1H 1832-33$^{\rm w}$ (NGC 6652)  & \cite{zan98b} \\
4U 1702-429                & \cite{cor03}	              & SLX 1744-300                     &                      & Ser X-1$^{\rm s}$                & \cite{cor02b} \\  
4U 1705-440                & \cite{cor03}	              & XB 1745-248 (Terzan 5)           &                      & 4U 1850-08           (NGC 6712)  &               \\
4U 1708-40                 &               & 4U 1746-37         (NGC 6441)    &                      & MXB 1906+000                     &               \\
\multicolumn{2}{l|}{1RXS J170854.4-321857$^{\rm a}$}           & GRS 1747-312$^{\rm w}$ (Terzan 6)& \cite{zan00a,zan03b} & Aql X-1                          &               \\
XTE J1709-267$^{\rm w}$    & \cite{coc98}  & EXO 1747-212                     &                      & 4U 1915-05                       &               \\  
XTE J1710-281              &     	   & GX 3+1$^{\rm s}$                 & \cite{ha03,cor03}         & XB 1940-04                       &               \\
2S 1711-339$^{\rm w}$      & \cite{cor02a} & SAX J1747.0-2853$^{\rm w}$       & \cite{we03,nat00,nat04}         & XTE J2123-058                    &               \\	
SAX J1712.6-3739$^{\rm w}$ & \cite{coc99c} & SAX J1748.9-2021$^{\rm w,n}$     & \cite{zan99b}        & 4U 2129 + 11           (M15)     & \cite{kuu02b}  \\	
MX 1716-31                 &               & SAX J1750.8-2900$^{\rm w}$       & \cite{nat99}         & 4U 2129+47                       &               \\	
RX J1718.4-4029$^{\rm w}$  & \cite{kap00}  & SAX J1752.4-3138$^{\rm w}$       & \cite{coc01}         & Cyg X-2                          &               \\	
XTE J1723-376         &                    & SAX J1753.5-2349$^{\rm w}$       & \cite{zan98a}        & SAX J2224.9+5421$^{\rm w,a}$     & \cite{cor02a} \\
                      &                    & XTE J1759-220                    &                      &                                  &               \\						     

\hline
\end{tabular}

$^{\rm w}$discovered through WFC observations;
$^{\rm a}$uncertain type-I classification; 
$^{\rm b}$underluminous burst from an underluminous source
         in a globular cluster \cite{got97};
$^{\rm n}$in NGC 6440;
$^{\rm s}$superburster; 
$^{\rm x}$Cen X-4 exhibited the brightest type-I X-ray bursts of all
bursters (25 Crab peak flux; \cite{mat80}).
\normalsize
\end{table*}

The list contains one burster which has not been published thus far:
1RXS J170854.4-321857.  Figure~\ref{figlc1708} presents the light
curve of the single burst-like event detected from this source with
the WFC in 1999. It exhibits a fast rise, a spectrum which is best fit
with a black body of temperature k$T=1$ to 2 keV (although other
models are formally acceptable as well), a 300~s duration, and its
1.9\arcmin-radius error circle (99\% confidence) is within 1.1\arcmin\
consistent with the ROSAT source 1RXS J170854.4-321857 which in the
mid 1990s exhibited an intensity of 5 to 20 mCrab in the ROSAT HRI
band and 0.3 mCrab in the PSPC band while its persistent emission was
never detected with the WFCs above an upper limit of 5~mCrab. The
double-peaked burst nature with a peak-to-peak delay of 3 minutes is
somewhat unusual for a type-I X-ray burst.

Because of the sheer volume of X-ray bursts, the WFCs picked up a
considerable number of bursts of rare kinds. Primarily these concern
the so-called 'superbursts', a phenomenon discovered with the WFCs
\cite{cor00}. These are X-ray bursts which last 10$^3$ times longer
than ordinary bursts and emit $10^3$ times more energy. After the
discovery, theoretical work quickly established that unstable carbon
burning \cite{woo76,str02} in a heavy-element ocean \cite{cum01},
possibly combined with photo-disintegration-triggered nuclear energy
release \cite{sch03}, is responsible for most superbursts. The carbon
is located in a deeper layer than the hydrogen and helium that is
burned in ordinary X-ray bursts, which accounts for the long duration
of the phenomenon. Currently, eight superbursts have been detected, 4
with the WFCs and 4 with instruments on RXTE.  The theory (e.g.,
\cite{cum03}) predicts that superbursts should occur on any neutron
star that is accreting matter at least as fast as one tenth of the
Eddington limit.  Thus far, they have only been seen in systems that
accrete near the lower boundary, but efforts are under way to search
for superbursts in more luminous neutron stars. The reason they were
not found thus far may be because more luminous systems tend to be
more variable in their persistent emission and provide less dynamic
range between the persistent flux level and the Eddington limit.  For
a more detailed review of (WFC results on) superbursts we refer to
\cite{ku03}.

Despite millions of seconds of coverage of many known LMXBs, 12 out of
the 54 bursters seen bursting by the WFCs exhibited just a single
burst. Many of those are low accretion rate LMXBs. Six had persistent
emission levels below the detection limit \cite{cor02a} and present an
interesting subset that is further discussed elsewhere in these
proceedings \cite{cor03b}.

\begin{figure}[t]
\includegraphics[width=\columnwidth]{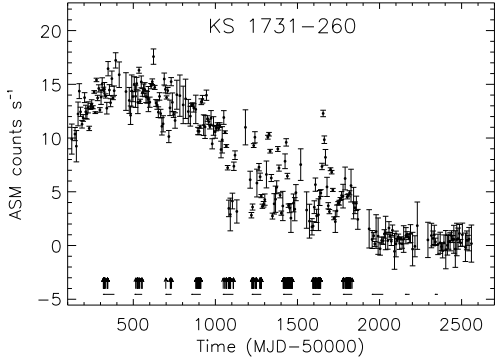}
\caption[]{KS 1731-260 2--12 keV light curve (from RXTE/ASM data)
and detections of bursts with the WFCs (arrows) during WFC visibility
periods (horizontal lines). From \cite{cor03}.\label{fig1731a}}
\end{figure}

\begin{figure}[!t]
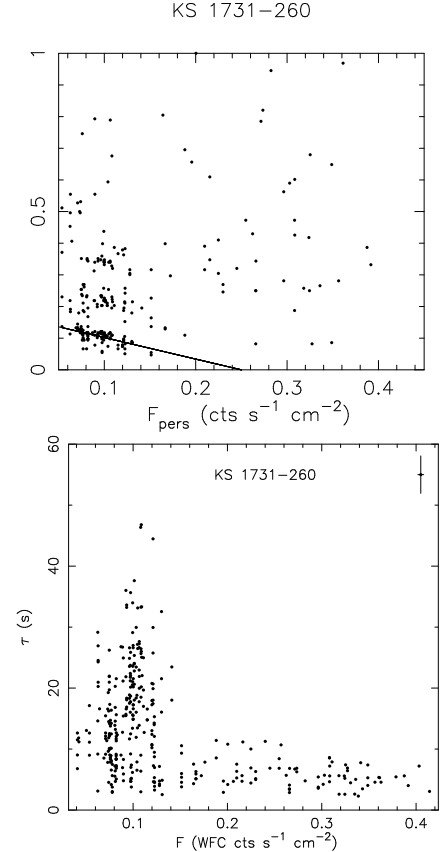

\centering
\includegraphics[angle=-90.,width=0.7\columnwidth]{intzand_f3a.ps}
\includegraphics[angle=-90.,width=0.75\columnwidth]{intzand_f3b.ps}
\caption[]{Dependencies of interval times between pairs (in days; top
panel) and e-folding decay time of bursts from KS 1731-260 (bottom
panel) on persistent flux. From \cite{cor03}.\label{fig1731b}}
\end{figure}

Cornelisse et al. \cite{cor03} investigated the burst properties of
the 9 most prolific bursters in the Galactic center region as a
function of mass accretion rate, using the complete WFC database.  The
number of bursts per burster ranges from 49 (for 4U~1820-30) to 423
(GX~354-0/4U~1728-33). The properties investigated are the wait time
from burst to burst, the burst duration and the presence of quasi
periodicities in burst occurrences.  With regards to the latter, an
intriguing discovery was made earlier by Ubertini et al. \cite{ube99}
with the WFCs: {\it all} bursts from GS~1826-24 between 1996 and 1998
recur after a quasi-fixed interval time of $5.76\pm0.26$~hr. Later the
coherence was found to be even higher because the period was seen
to have a decreasing trend that matched a gradual increase of the
persistent flux and, ergo, mass accretion rate \cite{coc99d} (see also
\cite{gal03}).

\begin{figure}[t]
\centering
\includegraphics[width=0.9\columnwidth]{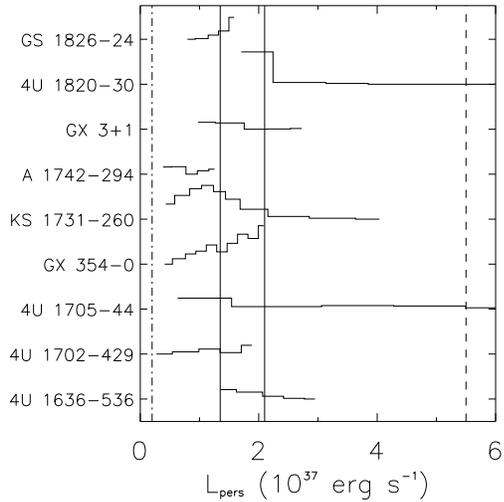}
\caption[]{A schematic diagram of the burst frequency as a function of
luminosity for the 9 most prolific bursters in the Galactic center
region. Between solid lines the rate mostly drops. No bursts were
detected to the right of the dashed line. The dash-dotted line
indicates the theoretical transition from hydrogen-rich to pure helium
bursts.  From \cite{cor03}.\label{figallen}}
\end{figure}

Particularly new findings uncovered by Cornelisse \cite{cor03} concern
KS~1731-260, because this is the only prolific burster which is
transient and, therefore, traces the largest dynamic range in mass
accretion rate. Figure~\ref{fig1731a} shows the light curve according
to data from the All-Sky Monitor (ASM) on RXTE, together with the
times that bursts were detected with the WFCs.  It is clear that the
burst rate increases dramatically when the transient starts to decay
(after having been active for 11 years). Actually this increase is
rather fast: the rate increases 5-fold while the persistent flux
decreases only 20\% from 0.15 to 0.12 WFC cts~s$^{-1}$cm$^{-2}$ (see
Fig.~\ref{fig1731b}). At the same time, the duration of the bursts
increases from a range of 2--12~s to 2--58~s and the burst recurrence
changes from random to quasi-periodic (see
Fig.~\ref{fig1731b}). Similar behavior has also been seen in most of
the other eight bursters investigated. GX~3+1 is another nice example
with a surprisingly sudden transition, see den Hartog et al.
\cite{ha03}.

The results are summarized in Fig.~\ref{figallen} which shows
schematically the burst rate as a function of luminosity. The trends
in all bursters are consistent with a universal behavior where the
burst rate increases from the lowest luminosities to roughly
1.3$\times10^{37}$~\lum\ and decreases above
2.1$\times10^{37}$~\lum. This is reminiscent of the three unstable
nuclear burning regimes (\cite{fuj81}, see also \cite{bil98}). In the
lowest regime, the accreted hydrogen is unstably burned and takes with
it the accreted helium. In the middle and upper regimes, the hydrogen
is continuously burned while the helium burns unstably in flashes. In
the middle regime a pure helium layer is formed which may be ignited;
in the upper regime also hydrogen is burned in the flash. Whenever
hydrogen is unstably burned (in the lowest and highest regimes),
bursts are prolonged by slow beta decays after proton captures by
heavy nuclei. Thus, bursts in the middle regime are always short
(i.e., shorter than 10~s). The boundary luminosities for the different
regimes depend on composition and temperature, but for solar
abundances they are predicted to be one order of magnitude lower than
observed. It is not clear why this is so (but see \cite{bil98}).

This interpretation of the burst rate versus luminosity relation
uncovers an inconsistency.  Bildsten \cite{bil00} presented GS~1826-24
as a textbook case of an X-ray burster in the {\em upper} burst
regime, based on an accurate measurement of the mass accretion rate
based on broadband BeppoSAX work \cite{zan99d} while the above work
suggests that it is in the {\em low} burst regime. It seems this
inconsistency can only be resolved by observations of this object at
higher mass accretion rates.  This will test whether the object will
start to show shorter (helium-fueled) and less frequent bursts like KS
1731-260 did in reverse (i.e., from higher to lower mass accretion
rates) which would prove it to be in the low burst regime. If so, we
would loose one of the only few textbook cases we have adhering to the
burst theory so nicely. Unfortunately, there are indications that the
brightening trend that GS 1826-24 exhibited since the launch of RXTE
and BeppoSAX (by about 40\%) is leveling off during the most recent
couple of years.

\section{LMXB TRANSIENTS}

The WFC observations revealed 14 previously unknown transients in the
Galactic center field that remained active for at least a few hours
(most for at least a few days; see Table~\ref{tab3}). Except perhaps
for two less obvious cases (SAX J1818.6-1703 \cite{zan98e} and IGR
J17544-2619\footnote{This object was not reported in the literature
since it was found in the data well after the outburst and was
suspected to be 1RXS J175428.3-262035. As observations with INTEGRAL
and XMM-Newton during a recent outburst in Sept. 2003 showed, the
transient and the ROSAT source are unrelated and WFC actually detected
a new transient \cite{sun03,gon03}}), they are all LMXBs. One contains
a dynamically confirmed black hole candidate (V4641 Sgr = SAX
J1819.3-2525; \cite{oro01}), one is suspected to be a black hole
candidate (SAX J1711.6-3808; \cite{zan02b}), seven contain neutron
stars (as established through X-ray burst detections), and one is
undetermined (SAX J1805.5-2031; \cite{low02}). Furthermore, the WFCs
detected recurrences of six already known transients. Three new
transients were detected only during type-I X-ray bursts; they are
so-called 'bursting-only sources' \cite{cor02a}. In total, the WFCs
detected 34 LMXB transients in the field, while the total number of
transients known to have been active during 1996-2002 is just two
larger: 36.

\begin{table*}[p]
\caption[]{36 LMXB transients that were seen to be active in 1996-2002
within 20 degrees from the galactic center. Many of the parameter
values for the bright transients were estimated from the publicly
available RXTE-ASM database (at URL {\tt http://xte.mit.edu}).  Other
characteristics were obtained from the references listed, that are
incomplete and emphasized on WFC results. This is an update from
\cite{zan01b}. Similar lists for transients before 1996 are given in
Chen et al. (1997).
\label{tab3}}
\begin{tabular}{lrrclll}
\hline
Name & Peak flux & Duration & Bursts? & Outburst Yrs & Comments & Refs.\\
     & mCrab     & days \\
\hline
GRO J1655-40     &  4500 & 200    &   & 94,95,96-97 & bhc             & \cite{kuu00} \\
X1658-298        &    30 & 1000   & y & 78, 99-01   & eclipses        & \cite{zan99c} \\
IGR J17091-3624  &    10 & $>10$  &   & 94,96,01,03 &                 & \cite{zan03c} \\
XTE J1709-267    &   200 & 100    & y & 97,02       &                 & \cite{coc98,jon03} \\
XTE J1710-281    &10     & $>1700$& y & 99-         & eclipses        & \cite{mar98,mar99} \\
SAX J1711.6-3808 &    60 & 170    &   & 01          & bhc             & \cite{zan02b} \\
2S 1711-339      &    50 &$700-1700$& y & 76, 98-     &                 & \cite{cor02a} \\
SAX J1712.6-3739 &    50 &$>$1100?& y & 98,01-      &                 & \cite{coc99c,zan02b} \\
RX J1718-4029    & \multicolumn{2}{c}{\em burst only}&y& [97]        && \cite{kap00} \\
XTE J1723-376    &    80 & 70-100 & y & 99              &                 & \cite{marshall99a,marshall99b} \\
Rapid burster    &   300 & 70     & y & every year      &                 & \cite{mas02} \\
GRS 1737-31      &    25 & 30     &   & 97              & bhc             & \cite{cui97} \\
GRS 1739-278    &   800 & 250    &   & 96              & jet, bhc        & \cite{var97} \\
XTE J1739-285    &   200 & 40     &   & 99,01,03        &                 & \cite{mar00a} \\ 
KS 1741-293      &    30 & few    & y & 89, 98          &                 & \cite{zan90,zan98c} \\
GRS 1741.9-2853  &    70 & 10?    & y & 90,96           &                 & \cite{coc99a}\\
XTE J1743-363    &    15 & $>600$ &   & 99-             &                 & \cite{mar99} \\
GRO J1744-28     &  2600 & 60     & y & 95-97           & type-II only?   & \\
EXO 1745-248     &   600 & $>200$ & y & 80,84,00,02     & any type-II?    & \cite{mar00b} \\
SAX J1747.0-2853 &   140 & 70     & y & 76?,98,99,00,01 &                 & \cite{we03,nat00,nat03} \\
GRS 1747-312     &    40 & 18     & y & every year      & eclipses        & \cite{zan00a,zan03b} \\
SAX J1748.9-2021 &    40 & 8      & y & 71?,98,01       &                 & \cite{zan99b,ver00,kaa03} \\
XTE J1748-288    &   500 &15      &   & 98              & jet, bhc        & \cite{rev00a}\\
SAX J1750.8-2900 &   120 & 230    & y & 97,01           &                 & \cite{nat99,kaa02}\\
XTE J1751-305    &    55 & 12     &   & 02              & ms pulsar       & \cite{mar02} \\
SAX J1752.3-3138 & \multicolumn{2}{c}{\em burst only}&y& [98]&            & \cite{coc01} \\
SAX J1753.5-2349 & \multicolumn{2}{c}{\em burst only}&y& [96]&            & \cite{zan98a}\\
IGR J17544-2619  &   200 & 0.1    &   & 96,97,99,00,03  & uncertain LMXB  & \cite{sun03,gon03} \\
XTE J1755-324    &   150 & 40     &   & 97              & bhc             & \cite{gol99} \\
2S 1803-245      &   700 & 25     & y & 76, 98          & jet             & \cite{mul98,rev00b}\\
SAX J1805.5-2031 &    50 & 250    &   & 02              &                 & \cite{low02} \\
SAX J1806.5-2215 &    13 & $>800$ &y  & 96              &                 & \cite{cor02a,zan98a} \\
SAX J1808.4-3658 &   100 & 18     & y & 96,98,00,02     & ms pulsar       & \cite{zan01}\\
SAX J1810.8-2609 &    15 &  3     & y & 98              &                 & \cite{nat01}\\
SAX J1818.6-1703 &   200 & 0.1    &   & 98              & uncertain LMXB  & \cite{zan98e}\\
SAX J1819.3-2525 & 12000 & 210    &   & 99,02,03        & jet, bhc        & \cite{zan00e,oro01} \\
\hline
\end{tabular}
\end{table*}

\begin{figure}[t]
\centering
\includegraphics[width=0.8\columnwidth]{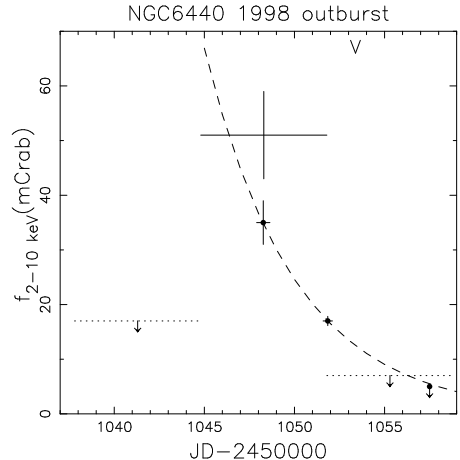}
\caption[]{The August 1998 outburst of the X-ray transient in
NGC\,6440 as observed with BeppoSAX ($\bullet$) and the RXTE ASM
(horizontal lines, solid for the detection, dotted for upper limits;
from \cite{ver00}). The dashed line indicates exponential decay with
e-folding time 5 days.  The V indicates the time of the optical
observations (see Fig.~\ref{figfrankopt}).
\label{figfrank}}
\end{figure}

When browsing through the list (Table~\ref{tab3}) it is striking that
there are only 8 out of the 36 transients that reached peak fluxes in
excess of 0.2 Crab units; a substantial fraction of 13 out of 36 is
even fainter than 0.05 Crab. Sometimes this is accompanied by a short
duration, but this is not the rule (e.g., \cite{zan01b}). An example
of this is given in Fig.~\ref{figfrank} which presents the light curve
of the 1998 outburst of SAX J1748.9-2021 in the globular cluster NGC
6440 which had an e-folding decay time as short as 5~d. A further
noteworthy detail is that most (22) transients burst and, thus,
contain a neutron star. The low outburst peak fluxes indicate
sub-Eddington luminosities for a canonical 8~kpc distance.

The fact that the observations have detected recurrence in half of all
transients suggests that a fair fraction of the LMXB transients with
recurrence times below a few years and on-times longer than a few days
have now been discovered in this field. Assuming that the sample is
complete and that this field represents half the total Galactic
population, there must be roughly 40 such transients in total. Quicker
transients, such as SAX~J1748.9-2021 \cite{ver00} or SAX~J1810.8-2609
\cite{nat01}, are more elusive unless they recur often enough (such as
V4641~Sgr, see Table~\ref{tab3}).

\section{X-RAY LUMINOUS OBJECTS IN GLOBULAR CLUSTERS}

There are 150 globular clusters in the Galaxy and 12 contain
13 luminous LMXBs (M15 has two). In a number of clusters,
considerable quantities of quiescent LMXBs have been found. Pooley et
al. \cite{poo03} estimate a total of 100 mostly quiescent LMXBs in
Galactic globular clusters. Thus, it is obvious that LMXBs are
abundant in clusters; for instance, luminous LMXBs are 100 times more
abundant than in the Galactic disk, in terms of number per unit mass
\cite{kat75,cla75}.

\begin{figure*}[t]
\centering
\includegraphics[width=\columnwidth,angle=270.]{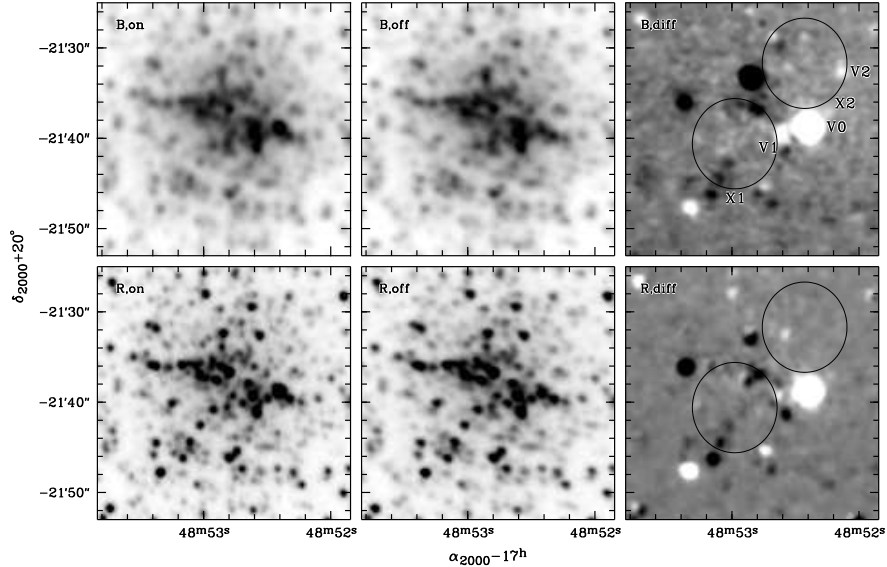}
\caption[]{B and R images of the field of NGC~6440, taken with SUSI at
the NTT when the X-ray source was still on, and with FORS1 on VLT-UT1
when it was off.  In these images, the brightest sources have
$B\simeq19$ and $R\simeq15$. Difference images are shown on the
right-hand side.  In these, stars that were brighter at the time the
X-ray source was on appear lighter, while those that were dimmer
appear darker.  The limiting magnitudes are $B\simeq24$ and
$R\simeq21$.  The error circles mark the positions of two X-ray
sources found in ROSAT HRI images.  The radii are 5$^{\prime\prime}$,
approximately equal to the boresight uncertainty in the ROSAT
positions. The optical counterpart to the luminous LMXB transient is
V2, as later confirmed by Chandra observations \cite{zan01c}. From
\cite{ver00}.
\label{figfrankopt}}
\end{figure*}

The nature of the compact accretor has been determined in 12 of
the 13 cases as neutron stars through the detection of type-I
X-ray bursts. The WFCs have established three of these bursters (see
Table~\ref{tab2}). The one unresolved case, AC211 in M15, has recently
been argued to contain a neutron star as well on the basis of an
optical study \cite{tor03}. Furthermore, AC211 is a persistent source
while all black hole LMXBs are transient.  Thus, another disk versus
cluster difference concerns the number ratio of black holes to neutron
stars in LMXBs. None of the luminous cluster LMXBs has a confirmed or
even a suspected black hole (0/13), while there are 37 cases in the
Galactic disk (of which 15 dynamically confirmed and 22 suspected;
\cite{mcc03}). The difference is not very significant though.  If the
same disk ratio of 37/150 applies to globular clusters as well, there
would be a 3\% probability that no black hole systems would be found
in a random sample of 13.

The WFCs made interesting observations of two particular luminous
globular cluster LMXBs. The first concerns NGC 6440. No X-ray outburst
was detected from this cluster since January 1972 \cite{mar75}, but in
August 1998 the WFCs detected an outburst which included first-time
observations of X-ray bursts. Follow-up observations were carried out
with the BeppoSAX-NFI and ROSAT, and Fig.~\ref{figfrank} shows the
combined light curve. The transient lasted only a little over a week,
and the peak flux was about 50~mCrab. Follow-up observations were also
carried out at high spatial resolution in the optical. Comparison with
later observations when the transient was off revealed the first-time
detection of an optical counterpart to an X-ray transient in a
globular cluster, see Fig.~\ref{figfrankopt} \cite{ver00}. This
identification was later confirmed by Chandra high-resolution X-ray
observations during another outburst \cite{zan01c}.  The fact that
some outbursts last very short makes clear that a swift response
of follow-up observations is sometimes imperative.

The other interesting WFC observation of a luminous cluster LMXB
concerns Terzan~6. First detected by Granat \cite{pav94} and ROSAT
\cite{ver95} in 1990, WFC and RXTE observations showed that this is a
transient with a very regular outburst pattern \cite{zan00a,zan03b}
like in the Rapid Burster and Aql X-1 with a (quasi) period of 4.5
months. By combining WFC data from the first half of the mission it
was found that this transient is one of now six known completely
eclipsing LMXBs in the Galaxy. The orbital period of 12.4~hr is
typical for an LMXB.

\section{CONCLUSION}

The results of the Galactic center campaigns of the BeppoSAX WFCs are
plentiful. This is mainly due to its large field of view which enabled
to monitor a large fraction of the Galactic LMXB population and
resulted in a large exposure of roughly 7 million seconds over six
years on tens of sources.  Such an exposure has not been accomplished
by any other device yet. Thus, it has been possible to detect rare
phenomena such as superbursts, burst-only sources and swift transients
(e.g., V4641 Sgr).  The analysis of WFC data will continue
and new results are expected to come out. This particularly concerns time scales
of a few hours and the longest time scales, and weak signals (e.g.,
X-ray bursts fainter than 0.5~Crab).

%

\section*{Acknowledgments}

The success of the Galactic center campaign was in part made possible
by the round-the-clock dedication of the duty scientists at the
BeppoSAX Science Operations Center in Rome and we thank them for their
efforts.  We acknowledge the ASI Science Data Center at Frascati and
SRON for continued support of BeppoSAX data analysis. We are grateful
to Rieks Jager and the team at SRON for building a wonderful
instrument.  The following institutes are acknowledged for financial
support: the Netherlands Organisation for Scientific Research, the
University of Utrecht, the Netherlands Research School for Astronomy,
CNR-IASF and the Italian Space Agency.


\begin{thebibliography}{9}

\bibitem{bl03}
J.A.M. Bleeker, these proceedings (2003)

\bibitem{jag97}
R. Jager et al, A\&AS 125 (1997) 557

\bibitem{pir03}
L. Piro \& L. Scarsi, these proceedings (2003)

\bibitem{hei03}
J. Heise et al., these proceedings (2003)

\bibitem{par95} J. van Paradijs,
	in W. Lewin, J. van Paradijs, E. van~den Heuvel (eds.), X-ray
  	Binaries, Cambridge University Press, Cambridge (1995), p.~536

\bibitem{ha03}
P. den Hartog et al., A\&A 400 (2003) 633

\bibitem{we03}
N. Werner et al., A\&A (2004) in press (astro-ph/0312377)

\bibitem{woo76}
S.E. Woosley \& R.E. Taam, Nature 263 (1976) 101

\bibitem{mar77}
L. Maraschi \& A. Cavaliere, Highlight of Astronomy 4 (1977) 127

\bibitem{le93}
W.H.G. Lewin et al., Space Sci.
        Rev. 62 (1993) 223

\bibitem{st03}
T.E. Strohmayer, L. \& Bildsten, in ``Compact Stellar
        X-Ray Sources, eds. W.H.G. Lewin and M. van der Klis, Cambridge
        University Press (2004) in press (astro-ph/0301544)

\bibitem{cum03}
A. Cumming, these proceedings (2003)

\bibitem{ku03}
E. Kuulkers, these proceedings (2003)

\bibitem{cor03b}
R. Cornelisse et al., these proceedings (2003)

\bibitem{gal03}
D.K. Galloway et al., ApJ 601 (2004) 466

\bibitem{liu01}
Q.Z. Liu et al., A\&A 368 (2001) 1021

\bibitem{ten86a}
A.F. Tennant, A.C. Fabian \& R.A. Shafer, MNRAS 219 (1986) 871

\bibitem{ten86b}
A.F. Tennant et al., MNRAS 221 (1986) P27

\bibitem{saz03}
P.M. Saz Parkinson et al., ApJ 595 (2003) 333

\bibitem{kuu02b}
E. Kuulkers et al., A\&A 399 (2002) 663

\bibitem{cor02a}
R. Cornelisse et al., A\&A 392 (2002) 885 

\bibitem{pir97}
L. Piro et al., IAUC 6538 (1997)

\bibitem{zan03a}
J.J.M. in 't Zand et al., A\&A 411 (2003) L487 

\bibitem{smi99}
M. Smith et al., in "The extreme universe" (3rd Integral workshop, Taormina, Sept.
    1998), Ap. Lett. Comm. 38 (1999) 137

\bibitem{cor03}
R. Cornelisse et al., A\&A 405 (2003) 1033 

\bibitem{zan99a}
J.J.M. in 't Zand et al., IAUC 7138 (1999)

\bibitem{coc98}
M. Cocchi et al., ApJ 508 (1998) L163 

\bibitem{coc99c}
M. Cocchi et al., IAUC 7247 (1999) 

\bibitem{kap00}
R. Kaptein et al., A\&A 358 (2000) L71 

\bibitem{coc99b}
M. Cocchi et al., in Proc. 5th Compton Symposium, 
    eds. M.L. McConnell and J.M. Ryan, AIP 510 (1999), p. 217 

\bibitem{kuu02}
E. Kuulkers et al., A\&A 382 (2002) 503  

\bibitem{baz97}
A. Bazzano et al., Proc. of
    the 4th Compton symposium, eds. C.D. Dermer, M.S. Strickman, J.D. Kurfess, AIP, (1997), p. 733

\bibitem{cor00} 
R. Cornelisse et al., A\&A 357 (2000) L21 

\bibitem{zan02a}
J.J.M. in 't Zand et al., A\&A 389 (2002) L43 

\bibitem{coc99a}
M. Cocchi et al., ApJ 523 (1999) L45 

\bibitem{zan90}
J.J.M. in 't Zand et al., Adv. Sp. Sc. 11 (1990) (8)187

\bibitem{zan98c}
J.J.M. in 't Zand et al., IAUC 6840 (1998)

\bibitem{zan00a}
J.J.M. in 't Zand et al., A\&A 355 (2000) 145 

\bibitem{zan03b}
J.J.M. in 't Zand et al., A\&A 406 (2003) 233 

\bibitem{nat00}
L. Natalucci et al., A\&A 543 (2000) L73 

\bibitem{nat04}
L. Natalucci et al., A\&A in press (2004) 

\bibitem{zan99b}
J.J.M. in 't Zand et al., A\&A 345 (1999) 100 

\bibitem{nat99}
L. Natalucci et al., A\&A 523 (1999) L45 

\bibitem{coc01}
M. Cocchi et al., A\&A 378 (2001) L37 

\bibitem{zan98a} 
J.J.M. in 't Zand et al., in "The active X-ray sky" (SAX/XTE conference, Rome 1997),
Nucl. Phys. B, 69 (1998) 228

\bibitem{mul98}
J.M. Muller et al., IAUC 6867 (1998) 

\bibitem{zan01}
J.J.M. in 't Zand et al., A\&A 372 (2001) 916 

\bibitem{nat01}
L. Natalucci et al., ApJ 536 (2001) 891 

\bibitem{coc00}
M. Cocchi et al., A\&A 357 (2000) 527 

\bibitem{ube99}
P. Ubertini et al., ApJ 514 (1999) L27 

\bibitem{coc99d} 
M. Cocchi et al., ibid, p. 203 

\bibitem{zan98b}
J.J.M. in 't Zand et al., A\&A 329 (1998) L37 

\bibitem{cor02b}
R. Cornelisse et al., A\&A 382 (2002) 174 

\bibitem{got97}
E.V. Gotthelf \& S.R. Kulkarni, ApJ 490 (1997) L161 

\bibitem{mat80}
M. Matsuoka et al., ApJ 240 (1980) L137

\bibitem{str02}
T.E. Strohmayer \& E. Brown, ApJ 566 (2002) 1042 

\bibitem{cum01}
A. Cumming \& L. Bildsten, ApJ 559 (2001) L127

\bibitem{sch03}
H. Schatz et al., ApJ 583 (2003) L90

\bibitem{fuj81}
M.Y. Fujimoto et al., ApJ 247 (1981) 267

\bibitem{bil98}
L. Bildsten, in 'The Many Faces of Neutron Stars', eds. J.
Buccheri, J. van Paradijs \& M.A. Alpar, Kluwer/Dordrecht (1998), p. 419

\bibitem{bil00} 
L. Bildsten, in "Cosmic Explosions", proceeding of the 10th Annual
October Astrophysics Conference (eds. S.S. Holt and W. W. Zhang), AIP
(2000), p. 359

\bibitem{zan99d}
J.J.M. in 't Zand et al., A\&A 347 (1999) 891 

\bibitem{kuu00}
E. Kuulkers et al., A\&A 358 (2000) 993 

\bibitem{zan99c}
J.J.M. in 't Zand et al., IAUC 7138 (1999)

\bibitem{zan03c}
J.J.M. in 't Zand et al., ATel 160 (2003) 

\bibitem{jon03}
P.G. Jonker et al., MNRAS 341 (2003) 823 

\bibitem{mar98}
C.B. Markwardt et al., IAUC 6998 (1998)

\bibitem{mar99}
C.B. Markwardt et al., IAUC 7120 (1999) 

\bibitem{zan02b}
J.J.M. in 't Zand et al., A\&A 390 (2002) 597 

\bibitem{marshall99a}
F.E. Marshall \& C.B. Markwardt, IAUC 7103 (1999) 

\bibitem{marshall99b}
F.E. Marshall et al., IAUC 7133 (1999) 

\bibitem{mas02}
N. Masetti, A\&A 381 (2002) L45

\bibitem{cui97}
W. Cui et al., ApJ 487 (1997) L73

\bibitem{var97}
M. Vargas et al., ApJ 476 (1997) L23

\bibitem{mar00a}
C.B. Markwardt et al., IAUC 7300 (2000)

\bibitem{mar00b}
C.B. Markwardt et al., IAUC 7482 (2000)

\bibitem{nat03}
L. Natalucci et al., A\&A (2004) in press 

\bibitem{ver00}
F. Verbunt et al., A\&A 359 (2000) 69

\bibitem{kaa03}
P. Kaaret et al., ApJ 598 (2003) 481

\bibitem{rev00a}
M. Revnivtsev et al., MNRAS 312 (2000) 151

\bibitem{kaa02}
P. Kaaret et al., ApJ 575 (2002) 1018 

\bibitem{mar02}
C.B. Markwardt et al., ApJ 575 (2002) L21

\bibitem{sun03}
R.A. Sunyaev et al., ATel 190 (2003)

\bibitem{gon03}
R. Gonzales-Riestra et al. A\&A in press (2004) (astro-ph/0402293)

\bibitem{gol99}
P. Goldoni et al., ApJ 511 (1999) 847

\bibitem{rev00b}
M.G. Revnivtsev et al., A\&A 344 (2000) L25

\bibitem{low02}
P. Lowes et al., IAUC 7843 (2002)

\bibitem{zan98e}
J.J.M. in 't Zand et al., IAUC 6840 (1998)

\bibitem{zan00e}
J.J.M. in 't Zand et al., A\&A 357 (2000) 520

\bibitem{oro01}
J.A. Orosz et al., ApJ 555 (2001) 489

\bibitem{zan01b}
J.J.M. in 't Zand, in "Exploring the gamma-ray universe" (4th INTEGRAL
workshop), eds. A. Gimenez, V. Reglero \& C. Winkler,
ESA SP-459 (2001), p. 463

\bibitem{poo03}
D. Pooley et al., ApJ 591 (2003) L131

\bibitem{kat75}
J.I. Katz, Nature 253, (1975) 981

\bibitem{cla75}
G.W. Clark, ApJ 199 (1975) L143

\bibitem{tor03}
M.A.P. Torres et al., MNRAS 341 (2003) 1231

\bibitem{mcc03}
J.E. McClintock \& R.A. Remillard, in ``Compact Stellar
        X-Ray Sources, eds. W.H.G. Lewin and M. van der Klis, Cambridge
        University Press (2004) in press (astro-ph/0306213)

\bibitem{mar75}
T.H. Markert et al., Nature 257 (1975) 32

\bibitem{zan01c}
J.J.M. in 't Zand et al., ApJ 563 (2001) L41

\bibitem{pav94}
M.N. Pavlinsky et al., ApJ 425 (1994) 110

\bibitem{ver95}
F. Verbunt et al., A\&A 300 (1995) 732

\end{thebibliography}
\end{document}